\documentclass{article}

\usepackage{PRIMEarxiv}

\usepackage[utf8]{inputenc} % allow utf-8 input
\usepackage[T1]{fontenc}    % use 8-bit T1 fonts
\usepackage{hyperref}       % hyperlinks
\usepackage{url}            % simple URL typesetting
\usepackage{booktabs}       % professional-quality tables
\usepackage{amsfonts}       % blackboard math symbols
\usepackage{nicefrac}       % compact symbols for 1/2, etc.
\usepackage{microtype}      % microtypography
\usepackage{lipsum}
\usepackage{fancyhdr}       % header
\usepackage{graphicx}       % graphics
\graphicspath{{media/}}     % organize your images and other figures under media/ folder

\title{Factors Hindering the Adoption of DevOps in the Saudi Software Industry}

\author{
  Mamdouh Alenezi \\
  College of Computer and Information Sciences \\
  Prince Sultan University \\
  Riyadh, Saudi Arabia\\
  \texttt{malenezi@psu.edu.sa}
}

\begin{document}
\maketitle

\begin{abstract}
DevOps has gained high importance in the global software industry due to the ease of software development, testing and deployment it provides. However, the Saudi software industry has not been able to adopt DevOps at a great pace due to various factors. This study, thus, aimed to examine different factors that hindered the adoption of DevOps in the Saudi software industry. Also, recommendations are provided at the end for Saudi Arabia to enhance the adoption of DevOps in its software industry. To accomplish the aims, this study used a literature review and interviews to gather data and examine it to produce the findings. The findings of the study highlight lack of support from organizational management and lack of laws as the major factors for the adoption of DevOps in the Saudi software industry.
\end{abstract}

% keywords can be removed

\section{Introduction}

The primary aim of the study is to evaluate the factors that hinder the adoption of DevOps in the Saudi software industry \cite{zarour2020devops}. The gap has been noticed by the software engineers that there is a lack of collaboration between operations and development teams in Saudi Arabia's software industry. The idea of collaboration has been considered an important aspect, and this creates cooperation between the two teams. The new process models called DevOps have been considered to discuss collaboration \cite{zarour2020devops}. Presently, IT organizations are not adopting DevOps as the added value or competitive gain, and thus, the companies are surviving a lot if they do not adopt it. The idea of the new process model aims to attain a fast and high-quality release of the software. DevOps is among the new software processes that help in the extension of the agility practices within the collaborative culture to empower the process of software delivery and development \cite{maroukian2020leading}.

The background reveals that the approach of DevOps is much concerned with improving the collaboration between the operation teams and development, which illustrates a new shift in the understanding of the way to build up the software system \cite{maroukian2020exploring}. It has also been denoted that the operational and development team have different goals in any project and the goal of the developer is to release the new aspect of the software. Thus, the goal of the operator is to keep the software stable and available. It is important to maintain the goals, and the coordination between the operational and development teams is considered vital in the project. The idea of DevOps helps to change the workflow of the traditional software development to streamline and accelerate the software delivery, which means to change the organizational culture in the delivery of software, not only the flow of process \cite{rafi2020towards}. 

The idea to adopt DevOps might spur the company to introduce new process changes, technological changes and personnel changes. The idea of software delivery has been treated as a continuous evolving procedure to obtain the expectations of users. DevOps makes it possible by bringing the operational and development teams together to facilitate the continuous integration, delivery and automation in the process that would result in reducing the time to market and would also result in enhancing the experience of customers and making the work quality-oriented \cite{rowse2021survey}. The research also depicts that there are more than 50\% of the companies across the industries, including manufacturing, banks and healthcare, are already involved in DevOps and have made it their digital strategy. The software DevOps has been prominent in academic and practical fields, but still, there is a lack of literature and academic publications regarding the empirical investigation of DevOps \cite{akbar2020identification}. 

The main motivation to evaluate the factors hindering the adoption of DevOps is to explore why organizations are still not able to adopt DevOps. The implementation of DevOps could increase the competitive advantage of the companies but go through the factors including lack of strategic direction, lack of knowledge, resistance to change, and silo mentality could help to understand why there is a lack of adoption of DevOps even in the large organizations. For this, the study focuses on this research question: "What are the factors hindering the adoption of DevOps in the Saudi Software Industry?"

In general, the following is needed to address most of the challenges of adopting DevOps. Software engineers need to know how to design and restructure their systems to integrate continuous delivery. Managers should focus on knowing DevOps and how it can be introduced and assessed as added value to the organization. Both managers and engineers should discuss and collaborate in choosing right level of automation toolset \cite{krey2022devops}. There are few Saudi IT companies that have adopted DevOps, but still, there are many who are facing hindrances to adopting it \cite{al2020devops}. Thus, the purpose of this study is to study the factors that are creating hindrances for Saudi IT companies in the adoption of DevOps.

\subsection{Google Trend (Saudi Arabia (DevOps))}

The google trend depicts the growth of DevOps in Saudi Software Industry from 2021 to 2022. The graph shows tremendous growth from 2021 to 2022, but there are fluctuations that have been making the graph unstable due to the hindrances that are making the companies difficult to adopt DevOps, especially in Eastern and Makkah Province.

\begin{figure}[htbp]
\centering
\fbox{\includegraphics[width=.92\linewidth]{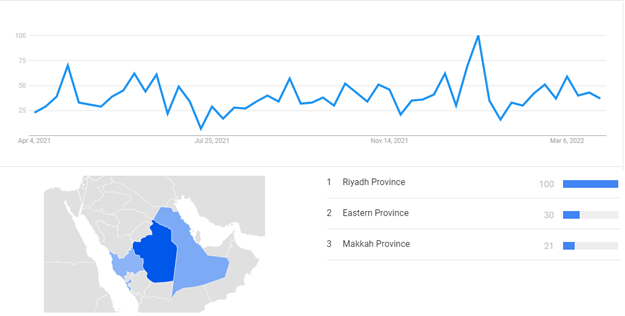}}
\caption{Google Trends \cite{google_2022}}
\label{fig:googletrends}
\end{figure}

\subsection{LinkedIn Insights (Saudi Arabia (DevOps)}

The LinkedIn insight shows that the companies in Saudi Arabia have been exploring DevOps engineers to implement it efficiently in their organization.

\begin{figure}[htbp]
\centering
\fbox{\includegraphics[width=.94\linewidth]{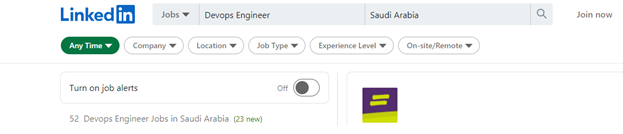}}
\caption{LinkedIn Insights  \cite{LinkedIn_2022}}
\label{fig:linkedin}
\end{figure}

\section{Literature Review}

\subsection{DevOps History}

Prior to the development of DevOps (Development and Operations) by Patrick Debois and Andrew Clay in 2008, the Agile methodology and waterfall model were used for the development of software \cite{perera2016evaluating}. The former deals with a linear approach of process execution with no option for testing errors in between the levels of processes, while the latter basis itself on rapid releases in new technologies, the collaboration of emerging processes and the introduction of improved customer feedback with linear process execution \cite{zarour2019research}. It releases the final product after the development of its constituent processes through the collaboration of stakeholders with room for continuous development in case of negative customer feedback. Agile did not focus on continuous development and integration of software with no room for monitoring of products (software) and customer feedback, and neither did it intend to work for the cultural aspect of organizations. With the introduction of DevOps and its development tools, agile techniques get integration with the elimination of their drawbacks and introduce a diversified culture in operations and development of software by integrating the efforts of all the team members ranging from developers in the deployment of the software, engineers in the testing of the software and administrators in documenting the products \cite{rowse2021survey}.

\subsection{DevOps Building Blocks}

DevOps integrate the cultural philosophies of an organization by merging all the departments in a single stream with the introduction of advanced practices based on some building blocks. It integrates the codes of different developers and stores them in a single repository for fellow developers to change and review in case of error \cite{laihonen2018adoption}. DevOps focuses on the correct metric and sets of measurements for the processes in the development and operations in the creation of software with continuous deliveries of tests and workings alongside customer feedback. Its iterative approach, tested from regular feedback in between the execution of processes, allow the formation of value-added products while eliminating practices that compromise quality and those rejected by the customers \cite{zarour2020devops}. This eliminates the gap between the developers and operation engineers through robust delivery and integration of service and the discard of time-consuming development processes. There are some processes in the system employed for the development of the software that needs regular check and balance, so in order to investigate the performance hindrance components, DevOps deploy a monitoring system in parallel with the development of the software to ensure optimized system performance \cite{erich2014report}. It also allows robust cycle-wise deployment and multiple scaling of the resources of the product that reduces time and cost by allowing regular change and updates in the system.

\subsection{DevOps Principles and Practices}

DevOps uses short feedback customer-related action loops that allow a continuous meeting of the standards proclaimed by the customers \cite{bou2017devops}. DevOps allows better communication and understanding between the employees in order to increase the productivity of the system as a whole. The workers are cross-skilled that can both develop and execute the tested codes. The workers understand the expertise, needs and functions of each and every other employee \cite{akbar2020identification}. It works in a culturally unified system where each and every member of all the departments know the actual processes going on during the development and operations of the products with a vertically integrated environment of full accountability for the performance and productivity of the system \cite{maroukian2021towards,ebert2016devops}. The developers, coders, engineers and administrators specialize in their respective skill sets and attributes and contribute towards a balanced environment with expertise in all domains. DevOps always allow room for continuous development in technologies, customer demands and organizational culture that work for cost optimization, time reduction and improved service of products. It also advocates the automation of industrial processes and information technologies \cite{erich2017qualitative}.

Nowadays, users and customers assume that modern software applications should accommodate their frequently changing requirements and needs \cite{gupta2017modeling}. To cope with these frequent changes, it is crucial for software companies to make frequents releases and deployments. TO achieve that, your environment has to be disciplined and transparent; otherwise, it may lead to major issues and failures including customer dissatisfaction. DevOps tries to close this gap and its guiding principles can be summed up as follows: culture, measurement, collaboration, and automation \cite{krey2022devops}. Gupta et al. \cite{gupta2017modeling} pinpointed four main factors that affect DevOps implementation namely, source control, automation, cohesive teams and continuous delivery.

With regards to DevOps cultural movement, there are usually four defining characteristics: open communication, incentive and responsibility alignment, respect, and trust. However, these cultural aspects cannot be the defining characteristics of DevOps but rather enablers to support software engineering process capabilities \cite{senapathi2018devops}.

\subsection{DevOps Practices Usage}

DevOps works for the development of high-end objectives, unlike traditional methods that cater to small-scale projects. DevOps involves continuous development in the automation and operations of the formation of software with the use of automation tools that allow bug detection and elimination from the codes \cite{alghamdi2019enhanced}. It works in a loop system where developers build code that is compiled by the team for testing by the operations team, which checks for the credibility of the code. It continuously plans and tests codes before dispatching them to production. The components of codes ready for production are rapidly identified and then sent for tracking of the usable parts of the codes under continuous integration with constant feedback between development and testing \cite{munoz2021guidance}. These tested codes are continuously being delivered to production. In case a new part of code needs to be incorporated in between the codes, DevOps uses a continuous deployment tool across different departments before providing codes for continuous monitoring that checks the code and system sending it. Further, the continuous monitoring system detects the problems in the processes while the codes are being dispatched for production. Thus, inaccuracies are removed, and efficiency and productivity are enhanced. The software is then released with an infrastructure code that allows configurations and changes during operations. These codes allow the automatic configuration of the software with respect to the environment in which the software is to be operated \cite{munoz2021guidance}.

DevOps is primarily used to enhance integration processes and allow fast solutions to problems resulting during production. With the advanced practices and unique principles of DevOps in the development and operation of software, widespread areas imply its applications \cite{masombuka2018devops}. DevOps shows significant results in financial trading companies where it automates the testing, building and development of the production processes with 45 second deployment time. It significantly reduces the time consumption of the processes with good customer feedback. Car production uses precise metrics and measurements of the scale of products. DevOps allows remedies for error correction and provides increased precision in error selection through its testing, monitoring and integration practices \cite{zarour2020devops}. Similarly, it shows a great importance in bug reduction while dealing with codes due to its robust code deployment in between operations. In networking industries, DevOps allows rapid installations of security patches for the protection of firewalls by the rapid introduction of solutions for bugs and viruses \cite{mansour2020proposal}.

\section{Research Methodology}

This paper uses two main techniques to examine the factors hindering the adoption of DevOps in the Saudi software industry. A literature review is conducted in this study to examine the technological, organizational, and environmental aspects that hinder the adoption of DevOps in the Saudi software industry. Semi-structured interviews were chosen as the main data collection method to get more insights about the current status of the Saudi software industry. The findings are presented in the following section.

Interviews have been selected to gather rich information about different experiences and perceptions of a small number of respondents \cite{boyce2006conducting, tech2018research}. In qualitative research, interviews are considered the most effective data collection method \cite{tech2018research}. Different questions were devised to collect the respondents' opinions about DevOps adoptions. The questions are Open-ended questions to give the freedom to the respondents to express themselves. 

DevOps practitioners from the operational and development teams of various Saudi software companies that implement the DevOps practices as a prerequisite are chosen as the participants of this research. These practitioners may function as different roles during different phases of development as a part of their work.

The following guidelines have been followed while conducting the interviews \cite{runeson2009guidelines}:
\begin{itemize}
  \item Clear interview purpose and how the data will be used.
  \item Introductory questions about background and role.
  \item Avoid personal/sensitive matters.
  \item Summarize the conclusions at the end of the interview.
   \item Record the interviews to transcribe them into text.
   \item Pre-schedule these interviews.
\end{itemize}

The following table \ref{tab:interview-questions} shows the interview questions used in this study. We tried to use a small set of questions in very clear English language to clarify and ease the process of collecting the data.

\begin{table}[htbp]
\centering
\caption{\bf Questions asked during the interviews}
\begin{tabular}{ll}
\hline
\# & Interview Question \\
\hline
1 & Could you please state your background - your name, role, and work experience? \\
2 & What does DevOps mean to you and your organization? \\
3 & What are the challenges of adopting DevOps practices? \\
4 & How do you go about solving those challenges? \\
5 & Which of those challenges are you able to successfully solve? \\
6 & What kind of support is needed to solve these challenges? \\
7 & Any other comments you would like to add? \\
\hline
\end{tabular}
  \label{tab:interview-questions}
\end{table}

\section{Results and Discussion}

This section discusses the impact of social, organizational, technological and environmental factors on the implementation of DevOps in an organization. Also, the findings are discussed based on their impact on the Saudi software industry.

\subsection{Interviews Details}

This section discusses the results of the interviews. A total of 12 participants working in different companies were selected.Table \ref{tab:interview-details} gives a description of these participants.

\begin{table}[htbp]
\centering
\caption{\bf Description of semi-structured interviewees}
\begin{tabular}{llll}
\hline
\# & Role & Experience & Company Size \\
\hline
1 & DevOps Engineer & 1-3 years & ~200 employees \\
2 & DevOps Architect & 3-5 years & ~300 employees \\
3 & DevOps Engineer & 3-5 years & ~160 employees \\
4 & Platform Engineer & 1-3 years & ~500 employees \\
5 & DevOps Engineer & 3-5 years & ~400 employees \\
6 & CI/CD Engineer & 3-5 years & ~120 employees \\
7 & Software Engineer & 1-3 years & ~400 employees \\
8 & Platform Engineer & 3-5 years & ~280 employees \\
9 & Software Engineer & 3-5 years & ~320 employees \\
10 & DevOps Engineer & 1-3 years & ~640 employees \\
11 & DevOps Architect & 3-5 years & ~480 employees \\
12 & DevOps Engineer & 1-3 years & ~240 employees \\

\hline
\end{tabular}
  \label{tab:interview-details}
\end{table}

\subsection{Technological aspect}

The technological strength of an organization defines the potential for introducing new technologies in an attempt to integrate them with the already established technologies of an organization. Large organizations, in contrast to smaller organizations, possess readily available technologies that account for the adaptation of DevOps in the software industry. The implementation of DevOps can significantly reduce costs for businesses in Saudi Arabia and has a significant impact on implementing DevOps in the software industry \cite{alghamdi2019enhanced}. Technological use of tools and processes accounts for a positive impact on DevOps adaptation. With the implementation of new technologies and the perceived benefits it reaps for the organization in terms of development and operations, businesses in Saudi Arabia significantly invest in adopting DevOps technologies due to highly anticipated positive outcomes. The integration of new operational services and technologies with the previous ones positively impacts the implementation of DevOps in Saudi Arabia \cite{munoz2021guidance}.

Security complications and concerns regarding its credibility promote threats to Saudi Arabia's software companies. Therefore, a strong and flawless security protocol with ample assurance from the security departments provides great motivation for officials from software companies in Saudi Arabia to invest in the introduction of DevOps. New technologies require severe testing, experimentation, and trials of the new processes and operations before putting them into real-life action. However, this factor has negligible importance for the adaptation of DevOps operations in Saudi Arabia's software companies \cite{rafi2021readiness}. 

\subsection{Organizational aspect}

Several organizational factors, like interest and support from top management levels of the organization, the scale and size of the company, knowledge of the processes and technologies implemented, and the eagerness of the management in the implementation of new processes, contribute towards the adaptation of DevOps in the software industry of Saudi Arabia \cite{masombuka2018devops}. Top officials and management in Saudi Arabia do not believe in the credibility and benefits of adopting DevOps technologies. The implementation of new technologies and advanced processes requires a large domain of operations and a considerably large size of the organization where the practices can be amply tested, experimented with and developed prior to their execution \cite{zarour2019research}. Saudi Arabia intends to adopt DevOps in large software organizations, and it already shows a greater induction of DevOps in big Saudi software industries than in smaller industries. It should not be neglected that the adaptation of new technologies is facile in smaller organizations, but when assessing the complexity of DevOps technologies, large organizations are preferred \cite{mansour2020proposal}.

\subsection{Environmental aspect}

Saudi Arabia experiences tough competition from software organizations, elevated pressures through government policies and laws, and a lack of trusted vendors of software products \cite{rafi2020rmdevops}. The technological advancements on national levels in Saudi Arabia are underdeveloped, with no local data centers and software houses. The competition between the software companies significantly defines the promotion and adaption of DevOps in Saudi Arabia, as Saudi Arabia already has few advancements in software industries, and the already established software companies provide tough competition in the development of operations to a considerable extent. In the previous research, the role of governmental legislation proved to be a hindrance in the development of the principles of DevOps in Saudi Arabia's industries \cite{hochbergs2020software}. In order to run software companies effectively with greater productivity, companies need to develop good relations with vendors. Saudi Arabia has a lack of vendor trust owing to a lack of data centers and developed technologies. A good vendor reputation accounts for the adaptation of DevOps principles in software companies. Saudi Arabia lacks a developed infrastructure of technologies with lower connection speeds owing to strict governmental policies and data transfer complications. This discourages software company owners from adapting to DevOps in their organizations as DevOps requires relatively high connection speeds with advanced technologies in the operations and development of software. However, research does not promote this fact as a hindrance to the development of DevOps in Saudi Arabia's software industry \cite{mohammad2018improve}. The effective development of DevOps in the software industry requires highly established data centers for the flow of data and information between different departments of the software industry and also with foreign data centers and companies. Saudi Arabia imposes strict policies regarding the transfer of data in and out of companies from their country. This considerably hinders the introduction of new technologies and DevOps principles. It is seen that local data centers provide a boost for new technologies and methodologies like DevOps \cite{maroukian2020exploring}.

DevOps is based on a strong, interlinked cultural environment in the software industry. It focuses on establishing a rich diversity of employees. Saudi Arabia imposes and exercises strict social values and culture. Assessing this factor, Saudi Arabia provides a hindrance to the adaptation of DevOps in the operations and processes of software companies. The growth of these companies is affected as a less diverse work environment culture contributes to a lack of communication between different departments within a company \cite{alenezi2019devops}.

\section{Recommendations}

Some of the recommendations for promoting the adaptation of DevOps in the Saudi software industry are as follows:

In a software company, there is significant automation in the processes of the DevOps method, ranging from development to testing and successful implementation of codes in the software industry \cite{hochbergs2020software}. Still, there is room for introducing more automation tools for testing new frameworks for the assessment of the correct framework for every specific industry in terms of their types and domains of operations \cite{mohammad2018improve}. Countries like Saudi Arabia sincerely need to introduce new laws and regulations regarding the introduction of new technologies like DevOps. They need to allow for the diversification of cultures in their software companies along with revised legislation to develop local software houses and data centers for an efficient flow of data. The development of DevOps needs proper documentation of its processes and technologies for industrialists, company owners, and employees in order to promote their interest in the development of emerging technologies and to improve communication levels among cross-department employees \cite{maroukian2020exploring}. 

In addition to this, the problems faced by Saudi Arabia regarding the implementation of DevOps also require a collaborative environment with enthusiasm from the upper management. There is less research that talks about the role of leadership types \cite{erich2017qualitative}. Therefore, this domain needs research to develop leadership types in accordance with the scope, size, and type of industries. Also, most of the researchers provide analysis of a specific target department of an industry. The domain of analysis needs to be enlarged, incorporating different departments and levels of hierarchy. Researchers also need to provide multiple solutions for each problem so that industrial practitioners adopt different strategies and methods for optimizing their business processes. Training sessions for developers, engineers, staff, and managers provide an effective understanding of the proper implementation of DevOps in software industry processes \cite{munoz2021guidance}. There is no proper management of data in many industries that require proper protection and security of information \cite{jaatun2017devops}.

\section{Conclusion}

Continuous and productive development in the software industry needs innovation and the introduction of new, improved technologies. DevOps is the latest emerging technological method that reaps great results for the optimization of industrial processes. This paper highlights all the challenges modern industries face while coping with increased modernization in software industries. This research paper provides a comprehensive analysis of the factors that promote and degrade the adaptation of emerging technologies and methods of industrial optimization and automation like DevOps and highlights the key factors that specifically hinder the adaptation of DevOps in the software industries of Saudi Arabia. This paper incorporates literature review and interviews. It explains the technological, social, organizational, and environmental factors that affect the implementation and adaptation of DevOps principles in Saudi Arabia. The paper covers extensive analysis of DevOps operations, including their processes, tools, and associated technologies. It especially highlights the key areas and domains for the Saudi Arabian software industry where improvement is anticipated in the future regarding the introduction of new industrial technologies and processes. It also highlights and recommends the key areas that require further research to come up with new processes and experiments.

%Bibliography
\bibliographystyle{unsrt}  
\bibliography{references}

\end{document}